\documentclass{aa}
 \usepackage{times}
 \usepackage{graphicx}
 \usepackage{psfig}
 \usepackage{amsmath}
 \usepackage{amssymb}

 \def\asec{\ifmmode ^{\prime\prime}\else$^{\prime\prime}$\fi}
 \def\lea{\ifmmode ^{<}_{\sim} \else $^{^{<}_{\sim}}$\fi}
 \def\gea{\ifmmode ^{>}_{\sim} \else $^{^{>}_{\sim}}$\fi}
 \def\lapp{\ifmmode\stackrel{<}{_{\sim}}\else$\stackrel{<}{_{\sim}}$\fi}
 \def\gapp{\ifmmode\stackrel{>}{_{\sim}}\else$\stackrel{>}{_{\sim}}$\fi}
 \def\HII{H\,{\sc ii}}
 
\begin{document}

 \thesaurus{06.                
	       (13.07.1;       
		12.03.3;       
		11.16.1;       
		11.19.3)       
	       }
 \title{Properties of the host galaxy of the gamma-ray burst
 970508 and local star-forming galaxies}
 \author{V. V. Sokolov\inst{1}, S. V. Zharikov\inst{1},
 Yu. V. Baryshev\inst{2},
 M. O. Hanski\inst{3}, K. Nilsson\inst{3}, P. Teerikorpi\inst{3},
 L. Nicastro\inst{4},
 and E. Palazzi\inst{5}}
     \offprints{ V. Sokolov}

     \institute{Special Astrophysical Observatory of R.A.S.,
  Karachai-Cherkessia, Nizhnij Arkhyz, 357147 Russia; sokolov,zhar@sao.ru
	 \and
  Astronomical Institute of St.Petersburg University, St.Petersburg
 198904, Russia; yuba@aispbu.spb.su
	 \and
  Tuorla Observatory, University of Turku, FIN-21500 Piikki\"o, Finland;
 mihanski,kani,pekkatee@astro.utu.fi
	   \and
 Istituto di Fisica Cosmica con Applicazioni all'Informatica, CNR,
  Via U. La Malfa 153, I-90146 Palermo, Italy
	   \and
 Istituto Tecnologie e Studio Radiazione Extraterrestre, CNR, Via Gobetti 101,
  I-40129 Bologna, Italy
	   }

  \date{Received / Accepted }

 \authorrunning{V. V. Sokolov et al.}
  \titlerunning{Host galaxy of GRB970508}

     \maketitle

     \begin{abstract}

 Late-time observations of GRB970508 with the SAO--RAS 6-m telescope
 in July--August 1998 show that the optical transient (OT)
 varies very little since November 1997.
 Here we report
 the behaviour of the $BVR_cI_c$ light-curves
 up to 470 days from
 the GRB occurrence.
 After $\simeq 200$ days any power-law decay has ceased and the OT
 contamination to the host galaxy flux is already less 
 than the observational errors.
 We derive the broad band spectrum of the host galaxy, without OT, and
 compare it with the average continuum spectra of galaxies of different
 Hubble types.
 The spectral  distribution of normal galaxies 
 with types earlier than Sbc 
 are confidently excluded.
 For $H_0 = 60$ km s$^{-1}$ Mpc$^{-1}$ and three Friedmann models
 with matter density and cosmological constant parameters $(\Omega_m,
 \Omega_{\Lambda}) =$ (1,0); (0,0); (0,1)  the derived host galaxy absolute
 magnitude  ($M_{B_{rest}}$) are  $-18.0\pm0.3$, $-18.5\pm0.3$ and
 $-19.0\pm0.3$ respectively.  
  The absolute $B$ magnitude $\approx -18.5$ corresponds to the luminosity
  $L_B\approx 0.1 L^*_{\rm gen}$ where $L^*_{\rm gen}$
   is the knee of the local general Schechter luminosity function.
   However  the luminosity of the GRB host is roughly  at the knee of the local
   luminosity function $L^*_{\rm late}$
   for late type galaxies Sd-Sm-Irr (Binggeli et al. 1988).
  Comparison of  the $BVR_cI_c$ spectral energy distribution
  of the GRB host with local starburst galaxies   
 leads to best fits for 
  the Scd starburst \HII\ galaxy NGC7793 and the blue
  compact galaxy Mrk 1267.  
  Position of the host galaxy in the $M_B$ vs. $\log D_{25}$ diagram
   for local late-type LEDA (Lyon-Meudon Extragalactic database) galaxies 
   allows us to attribute  
   GRB970508 host galaxy  
   to the blue compact galaxies.  

  \keywords{Gamma rays: bursts --- Cosmology: observation ---
 Galaxies: photometry -- starbursts}

  \end{abstract}

 \section{Introduction}

 At the present time, eight optical counterparts for gamma-ray bursts
(GRB) are known, all of them after detection with
 the Wide Field Cameras on board the Italian-Dutch
 BeppoSAX observatory.  They are:
 GRB 970228 (Groot et al. 1997; Van Paradijs et al. 1997),
 GRB 970508 (Bond 1997),
 GRB 971214 (Halpern et al. 1997),
 GRB 980326 (Groot et al. 1998; Eichelberger et al. 1998),
 GRB 980329 (Djorgovski et al. 1998),
 GRB 980519 (Jaunsen et al. 1998),
 GRB 980613 (Hjorth et al. 1998),
 GRB 980703 (Frail at al. 1998; Zapatero-Osorio et al. 1998).
 Galama et al. (1998a,b)
 reported about a possible supernova optical candidate for the GRB 980425.
 But a connection
 between type-Ib SN and the GRB (Pian et al. 1998) is still unclear.
 In all but GRB 980613 it is possible to speak about the detection of an
 underlying host galaxy (Hogg \& Fruchter 1998).

 This paper is dedicated to a detailed analysis and interpretation
 of new $BVR_cI_c$ photometrical data of the GRB970508
 host galaxy.
 Since our  $BVR_cI_c$ observations of the optical source related
 to GRB970508 in November 1997, we have continued them in
 January, July and August 1998. These late-time observations were
 undertaken with the main purpose to check how constant
 this faint ($R_c\approx 25$) source is up to 470 days after the
 gamma-ray burst, i.e. when the contribution of the OT
 can be considered to be comparable with or even
 less than the observational errors of brightness estimates
 in all $BVR_cI_c$ photometrical bands.
 These observations do show that  the optical counterpart
 varies very little since November 1997
 and allow us to determine the apparent magnitudes of the host galaxy.
 In this paper we compared the $BVR_cI_c$ spectrum of the GRB970508 
 host galaxy with spectral energy distributions of normal galaxies of 
 different Hubble types and extended the comparison to include local 
 star-forming galaxies. 

 \section{The GRB970508 host galaxy -- $BVR_cI_c$ late-time observations}

 Table \ref{tab1} presents the results of our observations
 with  the 6-m telescope of SAO--RAS in November ($I_c$ band) and in
  December 1997 ($B$ band), in January, July and August 1998
 ($BVR_cI_c$ bands) in one photometric system.
 The photometrical analysis and details are reported in Sokolov et al. (1998).
 The $B$ point in Dec. 1997 and $B$, $R_c$ points in Aug. 1998 were obtained
 by A. I. Kopylov.

 \begin{table}
 \caption[]{Observations summary ($t_o$ = 8.904 May 1997).}
 \begin{tabular}{lclcl}
 \hline
 Date UT & $t-t_o$ & Band & Time$_{exp}$ & \multicolumn{1}{c}{Magnitude} \\
         & (days)  &      & (s) & \\
 \hline
 25.00 Nov. 1997& 201.09 & $I_c$ & 4800    & $23.90\pm 0.14$ \\
 01.00 Dec.     & 206.01 & $B$   & 2400    & $25.75\pm 0.30$ \\
 24.87 Jan. 1998& 260.96 & $R_c$ & 2580    & $24.96\pm 0.17$ \\
 24.87 Jan.     & 260.96 & $V$   & 2400    & $25.44\pm 0.25$ \\
 22.99 Jul.     & 440.08 & $R_c$ & 3000    & $24.90\pm 0.16$ \\
 23.95 Jul.     & 441.05 & $B$   & 2400    & $25.80\pm 0.30$ \\
 23.95 Jul.     & 441.05 & $V$   & 2000    & $25.25\pm 0.22$ \\
 23.95 Jul.     & 441.05 & $I_c$ & 2000    & $24.07\pm 0.25$ \\
 21.74 Aug.     & 469.84 & $B$   & 4200    & $25.77\pm 0.19$ \\
 21.74 Aug      & 469.84 & $R_c$ & 3000    & $24.99\pm 0.17$ \\ \hline
\label{tab1}
 \end{tabular}
 \end{table}

 If $BVR_cI_c$ magnitudes of the optical counterpart did not change
 during the last half-year we can conclude that we observe a pure
 host galaxy without the optical remnant of GRB970508.
 Corresponding stellar magnitudes and fluxes are given in Table \ref{tab2}, 
 variant (or case) 1.  
 On the other hand,  the brightness can change within the errors,
 and the brighness decay of the optical remnant can be still described by
 a power-law relation during about a year after the burst occurrence.
 In that case the flux can be determined by fitting
 the observed $BVR_cI_c$ light curves with a two-component model,
 a sum of the
 optical remnant of GRB970508,
 fading according to a power-law, and a constant brightness
 host galaxy (see Zharikov et al. (1998) for details):
 $ F = F_o\times t^{\alpha}+F_c$.

  \begin{figure*}
  \resizebox{\hsize}{!}{\includegraphics{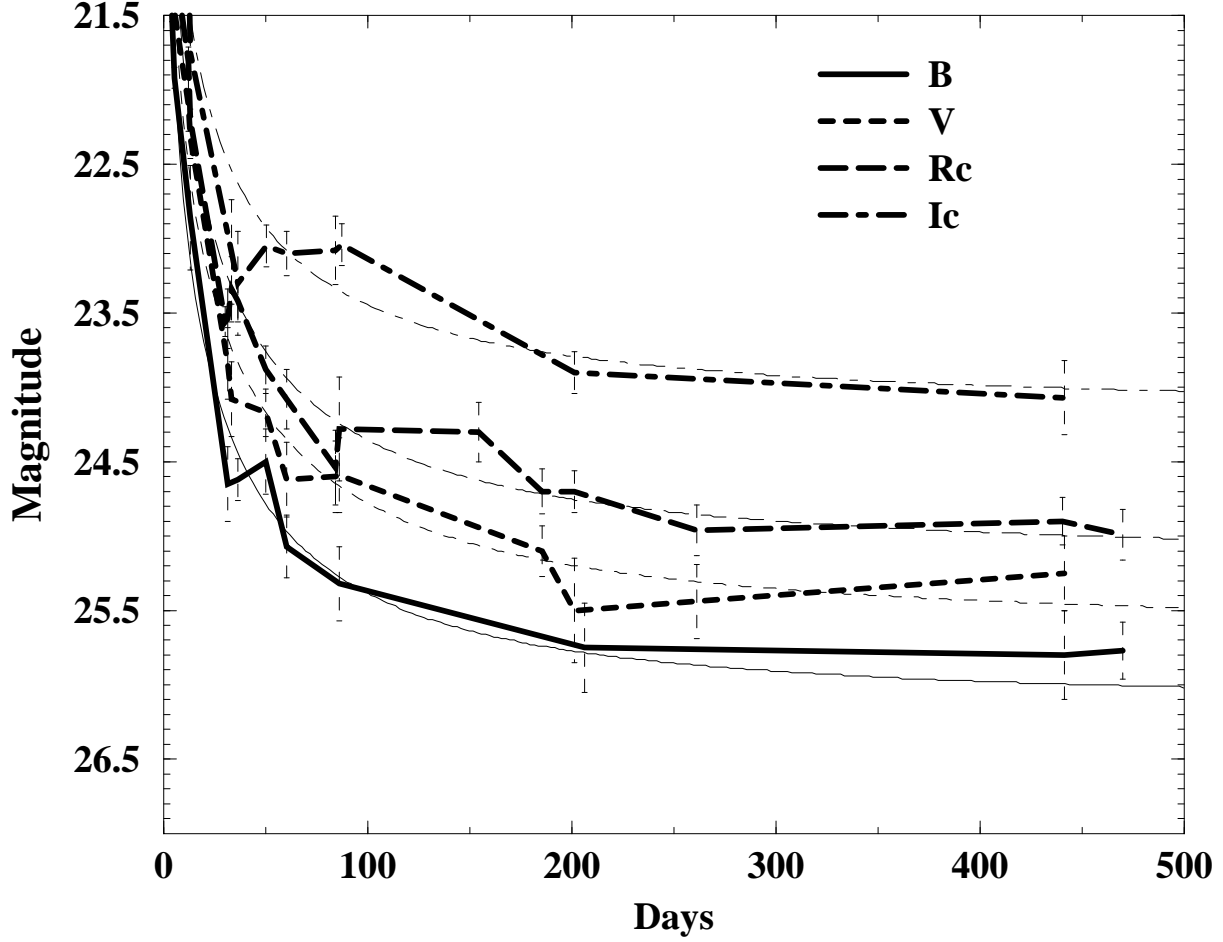}}
  \caption{
 The late $BVR_cI_c$ light-curves behavior of the OT + host galaxy of GRB970508
 up to $\simeq 470$ days from the time of the GRB. Four independent
 $BVR_cI_c$ power-law fits ($ F = F_o\times t^{\alpha}+F_c$, see Table 2)
 with different ${\alpha}$ are
 pictured by the thin lines.
 }
  \label{fig1}
  \end{figure*}

 To investigate the possible variability of a faint source we must
 avoid any systematic shifts in the observational data
 due to different photometric systems in various instruments.
 That is why for these $\chi^2$ fits (Table 2.)
 we used the data from the 6-m telescope only
 (from 10.77 May 1977 UT to 21.74 Aug 1988 UT)  
 unlike what was shown in the paper by Zharikov et al. (1998) 
 for $R_c$ and $B$ filters.
 So, the four independent $BVR_cI_c$ power-law fits
 with different slopes (Fig. \ref{fig1})
 give magnitudes for the host galaxy reported
 in Table \ref{tab2} variant 2.
 In comparison,
 an average power-law slope with
 $<\alpha>=-1.25\pm0.05$ gives the following magnitudes
 for the constant source (see Table \ref{tab2}, variant 3).

\begin{figure}
  \resizebox{\hsize}{!}{\includegraphics{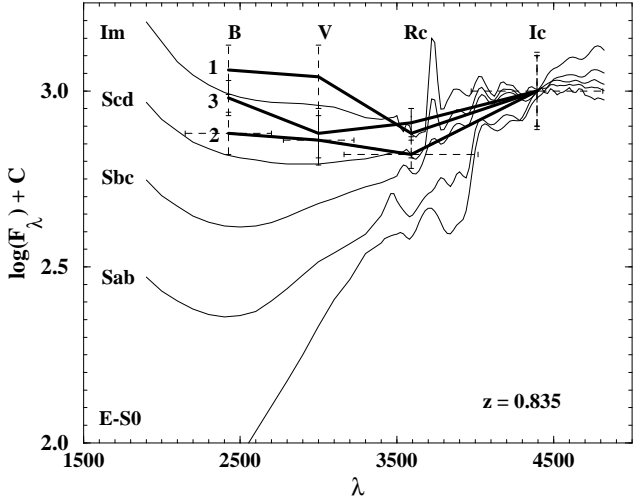}}
  \caption{
 A comparison of GRB970508 broad-band
  rest-frame ($z = 0.835$)
   spectrum
 log(F$_{\lambda}$) = logF$_{\lambda}$(obs)+C
 to average continuum spectra of galaxies of different Hubble types.
 No. 1, 2, 3 correspond to the three cases reported in Table 2.
 The spectra were shifted by some arbitrary constants for
 the best fits. F$_{\lambda}$ is in erg cm$^{-2}$ s$^{-1}$ \AA$^{-1}$ 
 and $\lambda$ is in \AA. FWHM of each filter for $\lambda_{eff} $
 with the account for $z = 0.835$ are denoted by dashed horizontal lines 
 with bars.
 }
\label{fig2}
  \end{figure}

 Accordingly, observational spectra in Fig.
 \ref{fig2}
 are presented for the 3 cases:
 the late-time observations (the case 1) and the two fits
 described above.
  In the cases 2 and 3 we indicate not observed, but
 ``theoretical" values and fluxes corresponding to $t$ tending
 to infinity.
 As can be seen from
 Table \ref{tab1}, the magnitude values in each individual band
 are almost constant during the period Nov. 97 -- Aug. 98, the last
 observations
 giving the  {\bf upper} estimates of the host brightness
 (spectrum 1 in Fig. \ref{fig2}).
 However if we
 force-fit a power-law brightness change
 then  {\bf lower} estimates are derived by fitting each observed band
 (spectrum 2 in Fig. \ref{fig2} ).
 Figure \ref{fig2} shows the broad-band spectrum of GRB970508 host galaxy
 compared with the typical spectra (Pence 1976) of different Hubble types 
 of galaxies normalized to the average value of $I_c$ flux 
 as measured for the host galaxy.

 So, we  choose the $BVR_cI_c$ spectrum of the host galaxy 
 as given in  Table \ref{tab2} (the variant 2) for
 interpretation in the following sections. However, all other
 variants in Table \ref{tab2} should be also kept in mind (as well
 as the observational errors)
 when comparing the results
 with the spectra of local galaxies.

 \begin{table}
 \caption[]{Magnitudes and fluxes for host galaxy of GRB970508.
1) The late-time observational magnitudes from Table 1.
2) $\chi^2$ fits: $F = F_o\times t^{\alpha}+F_c$.
3) The same $\chi^2$ fit for $<\alpha>=-1.25 \pm 0.05$.
The photometry with the BTA was reported by Sokolov et al. (1998) and 
Zharikov et al. (1998). 
}
 \begin{tabular}{lllllll}
\hline
 Band & \multicolumn{1}{c}{Magnitude} &
  \multicolumn{1}{c}{log F$_{\lambda,obs}$} & $\underline{\chi^2}$\\
 & & \multicolumn{1}{c}{
$\left(\frac{\rm erg}{{\rm cm}^2\ {\rm s}\ {\rm \AA}}\right)$}& (d.o.f) \\
 & & & \\ \hline
{\bf 1)} &&& \\
    $B$ & 25.77 $\pm$ 0.19   &  $-18.52 \pm 0.07$& \\
    $V$ & 25.25 $\pm$ 0.22   &  $-18.54 \pm 0.09$& \\
    $R_c$ & 24.99 $\pm$ 0.17 &  $-18.66 \pm 0.07$& \\
    $I_c$ & 24.07 $\pm$ 0.25 &  $-18.58 \pm 0.10$& \\ \hline
{\bf  2)} &&& \\
    $B$ ($\alpha=-1.32 \pm 0.05$)  & 25.99 $\pm$ 0.11 &  $-18.60 \pm 0.05$& 14.2/11   \\
    $V$ ($\alpha=-1.24 \pm 0.07$)  & 25.65 $\pm$ 0.17 &  $-18.70 \pm 0.07$& 36.6/14   \\
    $R_c$ ($\alpha=-1.25 \pm 0.04$) & 25.16 $\pm$ 0.09 &  $-18.73 \pm 0.04$& 52.7/19 \\
    $I_c$ ($\alpha=-1.18 \pm 0.07$)& 24.17 $\pm$ 0.28 &  $-18.62 \pm 0.11$& 44.3/12 \\ \hline
{\bf  3)} &&& \\
    $B$ & 26.15 $\pm$ 0.16   &  $-18.67 \pm 0.06$&18.9/12\\
    $V$ & 25.61 $\pm$ 0.16   &  $-18.69 \pm 0.07$&36.1/15\\
    $R_c$ & 25.16 $\pm$ 0.09 &  $-18.73 \pm 0.04$&52.7/18\\
    $I_c$ & 23.99 $\pm$ 0.25 &  $-18.55 \pm 0.10$&47.1/13\\ \hline
\label{tab2}
 \end{tabular}
 \end{table}

  \begin{figure}
  \resizebox{\hsize}{!}{\includegraphics{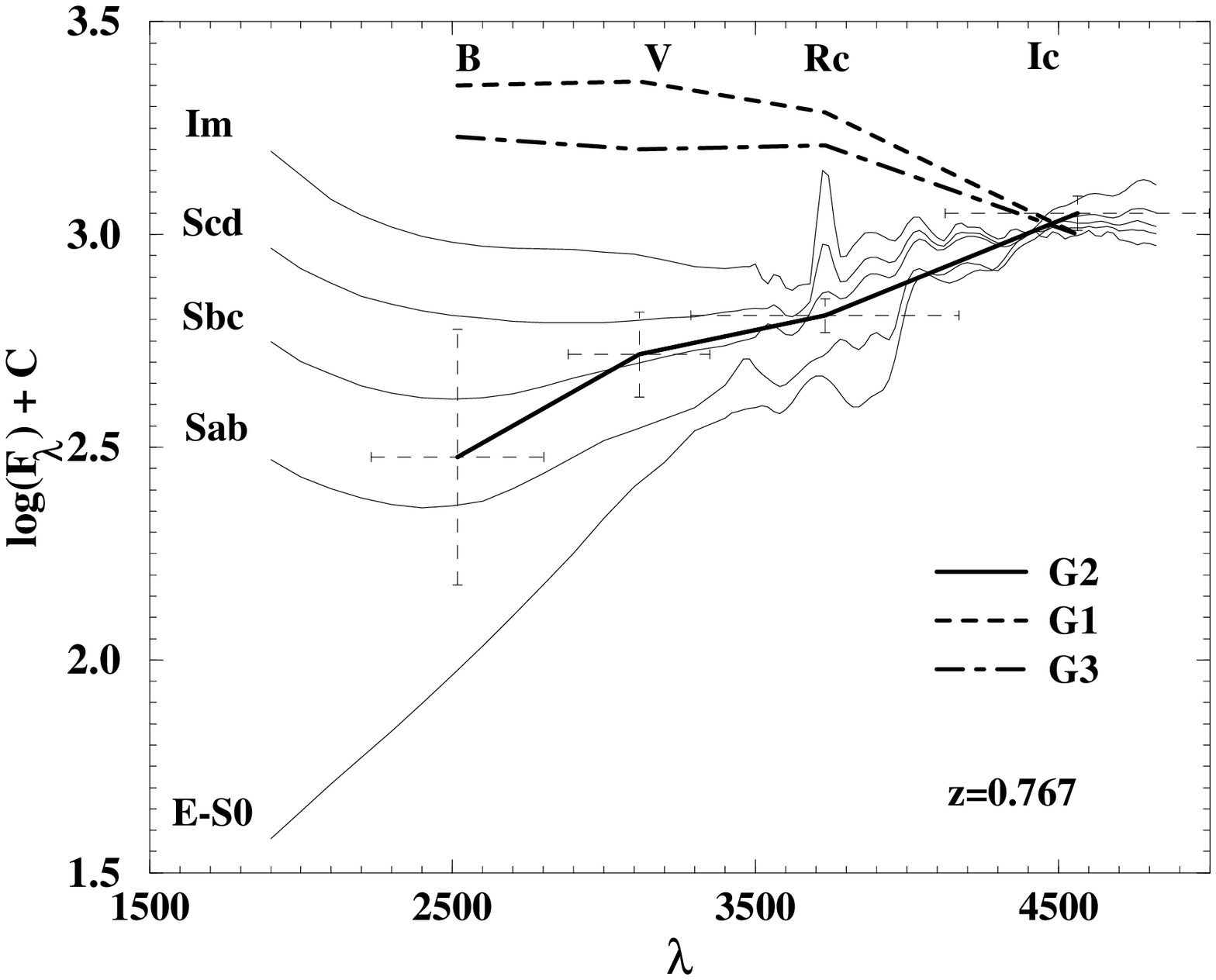}}
  \caption{
  Comparison of G2 (G1 and G3) broad-band 
  rest-frame spectra ($z = 0.767$) to
  typical average continuum spectra of galaxies.
 F$_{\lambda}$ is in erg cm$^{-2}$ s$^{-1}$ \AA$^{-1}$ and $\lambda$ is in \AA.
 See also the caption of Fig. 2. Magnitudes and fluxes for GRB host field
 objects were taken from Zharikov et al. (1998). 
 }
   \label{fig3}
  \end{figure}

 Zharikov et al. (1998) report a photometric study of the field of GRB970508.
 They performed $BVR_cI_c$ observations of the OT and of 3 nearby galaxies
 (named G1, G2 and G3, see their paper for more details).
 They conclude that only
 G2 may be responsible for the
 observed absorption system at redshift 0.767 (Metzger et al. 1997a).
 G1 and G3 are very blue galaxies with $z$ probably greater than 1,
 maybe in the range 1.5 -- 2.5.
 There are no optical lines for identifying redshifts
 in that $z$ range. So it is not
  surprising that the spectrum
 of the nearby galaxy G1 shows a relatively featureless, blue continuum
 (Bloom et al. 1998).    
 Figure \ref{fig3} shows the broad-band spectrum of the field galaxy G2 (G1 and G3)
 compared with
 those of typical galaxies of different Hubble types at
 $z = 0.767$.

  \section{Absolute magnitudes of the GRB970508 host galaxy}

  In order to compare the properties of the host galaxy with those
  of the local galaxy population, one needs to derive possible
  intervals for absolute magnitude,  linear size, and spectral energy
    distribution. To do this a cosmological model, redshift,
    and K-correction are required.

    The estimate of intrinsic physical parameters of extragalactic objects
  with redshifts approaching 1 depends rather sensitively on the
  adopted cosmological model.  The standard Friedmann model contains
    three parameters: Hubble constant $H_0$, matter density parameter
    $\Omega_m$, and cosmological constant parameter $\Omega_\Lambda$.
    Recent studies of the Hubble constant put it within the range 50 -- 70
    km s$^{-1}$ Mpc$^{-1}$ (see Theureau et al. 1997).
   In this paper we adopt $H_0 = 60$ km s$^{-1}$ Mpc$^{-1}$ .
   The values of $\Omega_m$ and $\Omega_\Lambda$ are
    observationally less constrained.  The recent work on the $m-z$
    test with supernovae of
    type Ia at redshifts up to 1 by Garnavich et al. (1998) make it
    imperative to consider in addition to the standard inflationary model
    also an empty universe with cosmological constant.
      For a review of
	  modern cosmological models and the needed mathematical
    relations, see Baryshev et
      al. (1994). Here we use  three Friedmann models which conveniently
      limit reasonable possibilities:
 \[ H_0 = 60 \mbox{ km s$^{-1}$ Mpc$^{-1}$, } \Omega_m = 1\mbox{, }\Omega_\Lambda = 0 \mbox{~~  (A)} \]
 \[ H_0 = 60 \mbox{ km s$^{-1}$ Mpc$^{-1}$, } \Omega_m = 0\mbox{, }\Omega_\Lambda = 0 \mbox{~~  (B)} \]
 \[ H_0 = 60 \mbox{ km s$^{-1}$ Mpc$^{-1}$, } \Omega_m = 0\mbox{, }\Omega_\Lambda = 1 \mbox{~~  (C)} \]
 For these models the relation $\Omega_m + \Omega_\Lambda
    + \Omega_k
 = 1$ is valid, where $\Omega_m = \rho_0 8 \pi G/3H_0^2$, $\Omega_\Lambda=
   \Lambda c^2/3H_0^2$, and $\Omega_k = -kc^2/R_0^2H_0^2$.  Here $\rho$,
 $\Lambda$, $k$, and $R$ are density, cosmological constant, curvature constant,
 and radius of curvature, respectively, and ``0" denotes the present epoch.
     The luminosity distance $R_{lum}$, the angular size distance $R_{ang}$
  and the proper metric distance $R_{p}$ are connected by the relation:
 \begin{equation}
 R_{lum} = R_{ang}(1+z)^2 = R_p(1+z)
 \label{equtwo}
 \end{equation}
 where the proper distances for the adopted models are given by
 \begin{equation}
 R_p=
 \begin{cases}
 R_H\: \frac{2(z-\sqrt{1+z}+1)}{1+z} & \text{ for model A,} \\
 R_H\: \frac{z(1+0.5z)}{1+z} & \text{  for model B,} \\
 R_H\: z& \text{ for model C.} \\
 \end{cases}
 \end{equation}
      Here $R_H = c/H_0$ is the present value of the Hubble radius. 
  The absolute magnitude $M_{(i)}$ of the source observed in filter ($i$) can
  be calculated from the magnitude-redshift relation
 \begin{equation}
 M_{(i)} = m_{(i)} - K_{(i)}(z) -5\log(R_{lum}/{\rm Mpc}) - 25
 \label{equthree}
 \end{equation}
  where $m_{(i)}$ is the observed magnitude of the object in the photometric
  band system ($i$) and $K_{(i)}(z)$ is the K-correction at redshift $z$,
    calculated from the rest-frame spectral energy distribution.

    The linear size $l$ of an object having an angular size $\theta$ is given
 by
   \begin{equation}
    l = \theta R_{ang} = \theta R_p /(1+z)
   \label{equfour}
   \end{equation}
 For $z = 0.835$, we get from Eq. (4) the linear sizes as
 normalized to $\theta = 1''$:
$l \approx 6.9 \text{ kpc } (\theta/1'') \text{ for model A}$,
$l \approx 8.5 \text{ kpc } (\theta/1'') \text{ for model B}$, 
and $l \approx 11.1 \text{ kpc } (\theta/1'') \text{ for model C}$. 

 A detailed discussion on the possible range of GRB970508 linear size
 is given in Sections 5 , after the discussion of the
 spectral energy distribution from our broad band
 photometry data, which can restrict the possible type of the host galaxy.

  The first spectral observations of the GRB970508 optical transient were
  obtained with the Keck-II 10-m telescope on 11 May 1997 (Metzger at al.
  1997a).  Two absorption line systems were detected: 1) $z = 0.835$ for
  the elements Fe\,{\sc ii}, Mg\,{\sc ii} and Mg\,{\sc i},
 2) $z = 0.767$ for Mg\,{\sc ii} doublet.  Such
  absorption features are commonly seen in the spectra of QSOs and are
  usually associated with galaxies along the line of sight to the QSO.
  This gives rise to two possibilities:
  1) we have detected the host galaxy with $z = 0.835$, or 2) we have detected
  a foreground galaxy on the line of sight, and the true host has $z >
  0.835$.
 Further optical follow-up observations, both imaging and spectroscopy,
 performed again with the Keck-II in June and November 1997 and February 1998,
 revealed an emission line [O\,{\sc ii}] corresponding to $z = 0.835$.
 The faint host galaxy of GRB970508
 seems to be responsible for this line
 (Metzger et al. 1997b; Bloom et al. 1998).

 For the calculation of the intrinsic parameters
  of the GRB host galaxy
 we adopt $BVR_cI_c$ magnitudes which correspond to
 case 2
 with none or minimal OT contamination (see Table 2).
The central $\lambda_{obs}$ = $\lambda_{eff}$ for our photometric system 
are equal correspondingly to: 
$\lambda_B = 4448$ \AA, $\lambda_V = 5505$ \AA, $\lambda_R = 6588$ \AA, $\lambda_I = 8060$ \AA.
(Corresponding FWHMs are equal to: 
$\Delta \lambda_B = 1008$ \AA, $\Delta \lambda_V = 827$ \AA, $\Delta \lambda_R = 1568$ \AA, 
$\Delta \lambda_I = 1542$ \AA.)  
 It is a useful coincidence that for z = 0.835 the observed wavelength of
  the $R_c$ filter corresponds to the rest band $U_{rest}$ ($\lambda_U
  = 3652$ \AA), and $I_c$ filter corresponds to the rest band $B_{rest}$
  ($\lambda_B = 4448$ \AA).
   By a simple shift $\lambda_{obs}$   to
	   $\lambda_{emit} = \lambda_{obs}/(1+z)$,
   in the $\log(F(\lambda_{obs}))$ vs.
 $\lambda_{obs}$ diagram,
we have four points of the rest frame continuum spectral  distribution,
which allow us to make a comparison with the available spectral distributions
 measured for local galaxies (see also Fig. 2 and Fig. 3).

   The K-correction in Eq. (3) can be calculated from the standard formula
   (Oke \& Sandage 1968):
 \begin{equation}
  K_{(i)}(z) = 2.5\log(1+z)
 +2.5\log\frac{\int_0^\infty F_{\lambda} S_{(i)}(\lambda) {\rm d}\lambda}
 {\int_0^\infty F_{\lambda/(1+z)} S_{(i)}(\lambda) {\rm d}\lambda}
 \label{equ5}
 \end{equation}
In this formula $F_{\lambda}$ is the rest-frame spectral energy
  distribution
 and $S_{(i)}(\lambda)$ is the transmission curve for
  filter (i).  
   As we mentioned above, the effective wavelength of the $I_c$ band
 for $z = 0.835$ happens to correspond to $B_{rest}$ band.  This allows
 us to calculate directly from Eq. (\ref{equ5}) the value of the K-correction
 for the $B$-magnitude, replacing the second term with
 $  2.5\log (F_{B_{rest}}/F_{\lambda_B /1+z}) \approx 2.5\log
 (F_{I_{obs}}/F_{B_{obs}})$.
  The  calculation yields
 \[ K_{B} = +0.6 \pm 0.3   \]
The absolute $B_{rest}$ magnitude from Eq. (3), (with K-correction
  $+0.6$ for $z = 0.835$) are:
$M_{B_{rest}} = -18.0 \pm 0.3$  for model A, 
$M_{B_{rest}} = -18.5 \pm 0.3$  for model B,
and $M_{B_{rest}} = -19.0 \pm 0.3$  for model C.

  We also estimate that  the K-correction for the field G2 galaxy at $z = 0.767$
 is $K_B$ $\approx 2.1$. 
 (Only an upper limit exists for the flux at $B$, see Fig. 2.).
  This allows us to derive $M_{B_{rest}}\lapp -18.4,\; -18.9,\; -19.4$
for the cosmological
 models A, B and C,  respectively.
 Following Natarajan et al. (1997) we can now calculate a lower
 limit to the maximum radius $R_{\rm max}$ of the gaseous halo of the galaxy G2
 which can produce the observed Mg\,{\sc ii} absorption line system:
 $R_{\rm max} \gapp 38$, 42 and 46 kpc (models A, B and C).
 The angular distance between the GRB host and G2 is
 about $4''$ which corresponds to a linear projected radius of 29,
 35 and 45 kpc, so we may conclude that the G2 galaxy could really be
 responsible for
 the observed absorption line Mg\,{\sc ii} at $z = 0.767$
 in the spectrum of the GRB970508 OT (Metzger et al. 1997a).

         \section{Comparison with ultraviolet spectra of local galaxies}

    In Fig. \ref{fig2} we compared the $BVR_cI_c$ spectrum of the GRB970508 host
    galaxy with the spectral energy distributions of normal galaxies of
    different Hubble types.  It appears difficult to match the
    observed spectrum with any single Hubble type.  However, for the
 GRB970508 host galaxy one may confidently reject early type galaxies
 from E to Sb.  Here we extend the comparison to include local star-forming
 galaxies.

\begin{figure}
  \resizebox{\hsize}{!}{\includegraphics{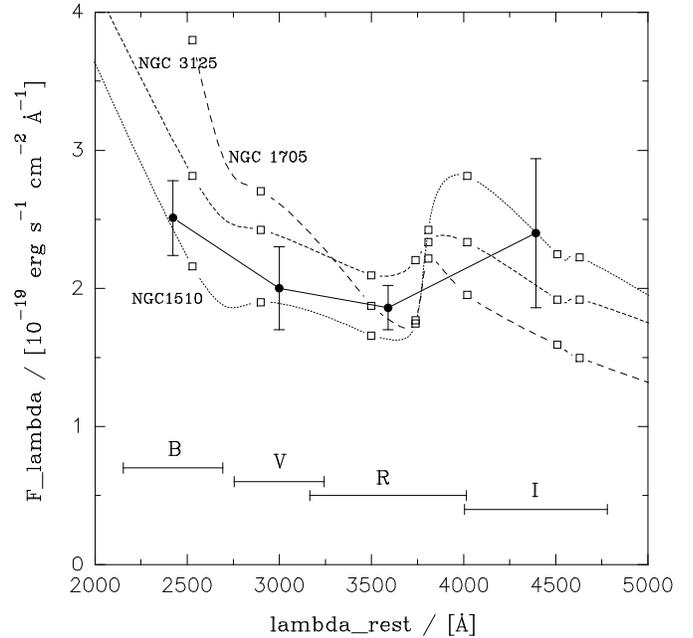}}
  \caption{
  A comparison of GRB970508  broad-band $BVR_cI_c$ spectrum    
  to continuum spectra F$_{\lambda}$
  of some blue compact dwarf galaxies
  in Kinney et al. (1993).
  The fluxes from Table \ref{tab2} (case 2) 
  have been scaled arbitrarily to the best fit. 
  The observing bands have been shifted to the rest frame ($z=0.835$) of
  GRB970508. Filled symbols indicate the observations and the smooth
  curves connect the selected continuum points, marked with open boxes.
  See text for details.
  }
  \label{bcdgfig}
  \end{figure}

\begin{figure}
  \resizebox{\hsize}{!}{\includegraphics{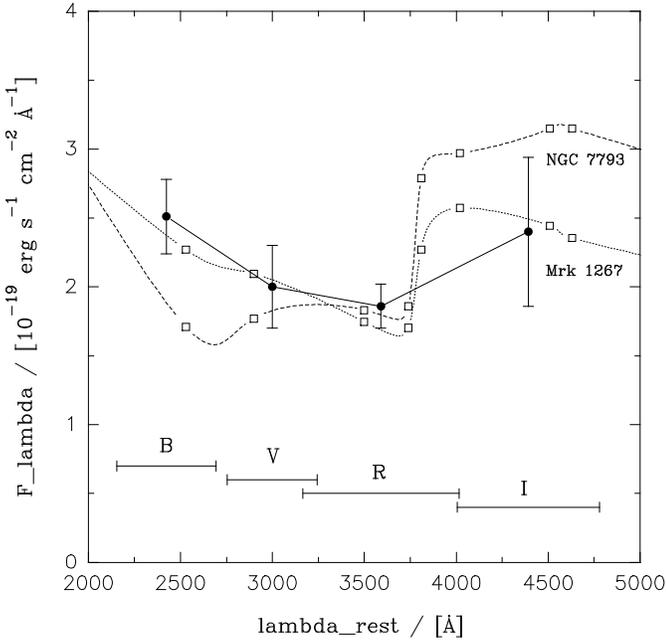}}
  \caption{
  A comparison of GRB970508 broad-band spectrum from Table \ref{tab2} (case 2) 
  to continuum spectra F$_{\lambda}$
  of a blue compact galaxy (Mrk 1267)
  and an Scd \HII\ galaxy (NGC7793). See also the caption of Fig. 4
  and text for details.
  }
  \label{bcghiifig}
  \end{figure}

  Figures \ref{bcdgfig} and \ref{bcghiifig} show a comparison between the
  continuum fluxes of the GRB970508 host (shifted to the rest frame) and
  continuum spectra of some star-forming galaxies  selected
  from the ultraviolet atlas of Kinney et al. (1993). We
  restricted  the search to the absolute magnitude interval $-21 <$
  $M_B < -17$ and looked for spectra that resemble the observed
  broadband spectrum of the GRB970508 host galaxy. For the latter task we
 used the tables in
  McQuade et al. (1995) and Storchi-Bergmann et al. (1995) who list
  continuum fluxes at selected wavelengths
  in the interval 1355 -- 7525 \AA. The catalog fluxes have been
  marked with open boxes in Figs. \ref{bcdgfig} and
  \ref{bcghiifig}. The fluxes have been scaled arbitrarily and smooth
  lines connect the points for better comparison with observed points.
The observed continuum points are reported with the associated error bars
 (a galactic extinction of A$_{\rm V} = 0.08$ was assumed
 (Djorgovski et al. 1997; Sokolov et al. 1998)).
 We have checked that changing the galactic extinction from a low value of
 0.01 to a higher value of 0.15 does not essentially alter the spectrum.

  All galaxies in Figs. \ref{bcdgfig} and \ref{bcghiifig} display
  also fairly strong emission lines, most notably at
  $\lambda=3727$ \AA\ [O\,{\sc ii}],
  $\lambda=4861$ \AA\ H$_{\beta}$, $\lambda = 4959,\; 5007$ \AA\ [O\,{\sc iii}]
  and $\lambda=6563$ \AA\ H$_{\alpha}$, which are caused by their star-forming
  activity. Of these strong lines only $\lambda=3727$ \AA\ [O\,{\sc ii}]
  falls within the range of our continuum observations, close to
  the $R_c$ band.

    We conclude that the best fits to the observed spectrum of the host
    galaxy are displayed by blue compact S0 dwarf galaxy NGC1510, the star-forming 
    \HII\ galaxy NGC7793 and
    the blue compact galaxy Mrk 1267.  The absolute $B$ magnitude of NGC7793
    ($-18.0$), which has the Hubble type Scd, is practically the same as
    that of the host, while Mrk 1267 is much brighter ($-20.8$) and its
    Hubble type is not well defined.

 \begin{table*}
 \caption[]{Parameters of the galaxies used in the comparison of the UV
 spectra, extracted from LEDA. Distances $r$ (with dispersions $\sigma_r$)
 are calculated using radial velocities corrected to Virgo infall and
 $H_0=60$  km~s\(^{-1} \)Mpc\(^{-1} \), except for NGC1705 (O'Connell et
 al. 1994) and NGC7793 (Puche \& Carignan 1988). 
 $B_{\rm T}$ are total apparent B-magnitudes,
 $\log{d_{25}}$ are log of 25 B-mag/arcsec$^2$ 
 engular isophote diameters in units of $0.1$ arcmin, 
 $D_{25}$ are linear 25 B-mag/arcsec$^2$ 
 isophote diameters with dispersions $\sigma_{D_{25}}$.
 }
\centering
  \begin{tabular}{llcrrrrrrrr}
 \hline
  PGC/LEDA id. & Alt.\ id. & Type & $\log{d_{25}}$ & $B_{\rm T}$ & $D_{25}$ &
 $\sigma_{D_{25}}$ & $M_B$ &  $\sigma_{M_B}$ & \multicolumn{1}{c}{$r$} &
 \multicolumn{1}{c}{$\sigma_r$} \\
	       &           &      &        &  & (kpc) & (kpc) &  &  &  (Mpc) & (Mpc) \\ \hline
   PGC 0014375 & {NGC    1510 }    & S0    &      1.429 &      13.51 & 9.37
  &       10.12 &     $-16.77$ &        .63 &      11.41 &        .14 \\
  PGC 0016282 & {NGC    1705 }     & E-S0  &      1.270 &      12.76 &     2.71
  &       1.11 &     $-15.73$ &        .87 &       5.0$^1$ &     2.0$^1$ \\
  PGC 0029366 & {NGC    3125 }     & E     &      1.047 &      13.47 &     5.59
  &        .85 &     $-17.38$ &        .31 &      14.81 &       1.51 \\
  PGC 0032672 & {MK 1267     } &       &       .644 &      14.16 &     12.31
  &       1.74 &     $-20.75$ &        .59 &      96.04 &        .72 \\
  PGC 0073049 & {NGC    7793 }     & Scd   &      1.982 &       9.70 &     9.43
  &       1.44 &     $-17.95$ &       .24 &       3.38$^2$ &      .30$^2$\\
  \end{tabular}
 \label{tab3}
  \end{table*}

           \section{Discussion}

    The presence of absorption lines as characteristic of star formation
 regions like Mg\,{\sc ii}, Mg\,{\sc i} and Fe\,{\sc ii} (Metzger et al. 1997a),
 evidence for emission line $\lambda=3727$ \AA\ [O\,{\sc ii}], and the
 similarity of the observed GRB host spectrum to the spectral energy
 distribution in a Scd galaxy prompts us to inquire how the host galaxy
 is might related to the class of late-type spiral and irregular galaxies.
    The first step in such a study is to compare the properties of the GRB 
    host galaxy with the galaxies in the local Universe, on which we
 have good knowledge.  The second step is to understand possible deviation
 from the local galaxy population (normal, starburst, irregular) as
    an effect of evolution and/or selection.

   The studies of the local luminosity function of types Sd - Sm - Irr
    (see e.g. Binggeli et al. 1988) indicate that these form a subclass
   among all galaxies with a knee in the luminosity function at about
 $M_B \approx -18,\; -19$. So, the luminosity of the GRB host is roughly  
 at the knee of the local luminosity function $L^*_{\rm late}$. 
 Though one should note that here the primary selection is not optical but
 comes from the gamma-ray burst,  
 this may explain why overluminous galaxies do
 not dominate among detected GRB hosts.

  In order to make further comparison between the GRB970508 host
  and local galaxies, especially to utilize the absolute
  magnitude--linear size relation ($M_B$ vs.\ $\log{D_{25}}$)  
  we used the Lyon-Meudon Extragalactic
  database (LEDA) which currently lists the
  main astrophysical parameters for more than 165\,000 galaxies.
  The data are collected from the major published galaxy surveys, and
  complemented with individual measurements.
  The parameters have been carefully reduced to a common system (of
  de Vaucouleurs et al. 1991) and corrected
  for any instrumental or observational imperfections (Paturel et al.
  1997).
  The latest information on the status of LEDA and the data itself
  are available at http://www-obs.univ-lyon1.fr/leda/\,.
  Extracted from LEDA
  parameters of the galaxies used in the comparison (Figs. 4 and 5)  
  with the GRB970508 host UV spectra are shown in  
  Table \ref{tab3}. 
  
\begin{figure*}[t]
  \includegraphics[angle=270,width=14.5cm]{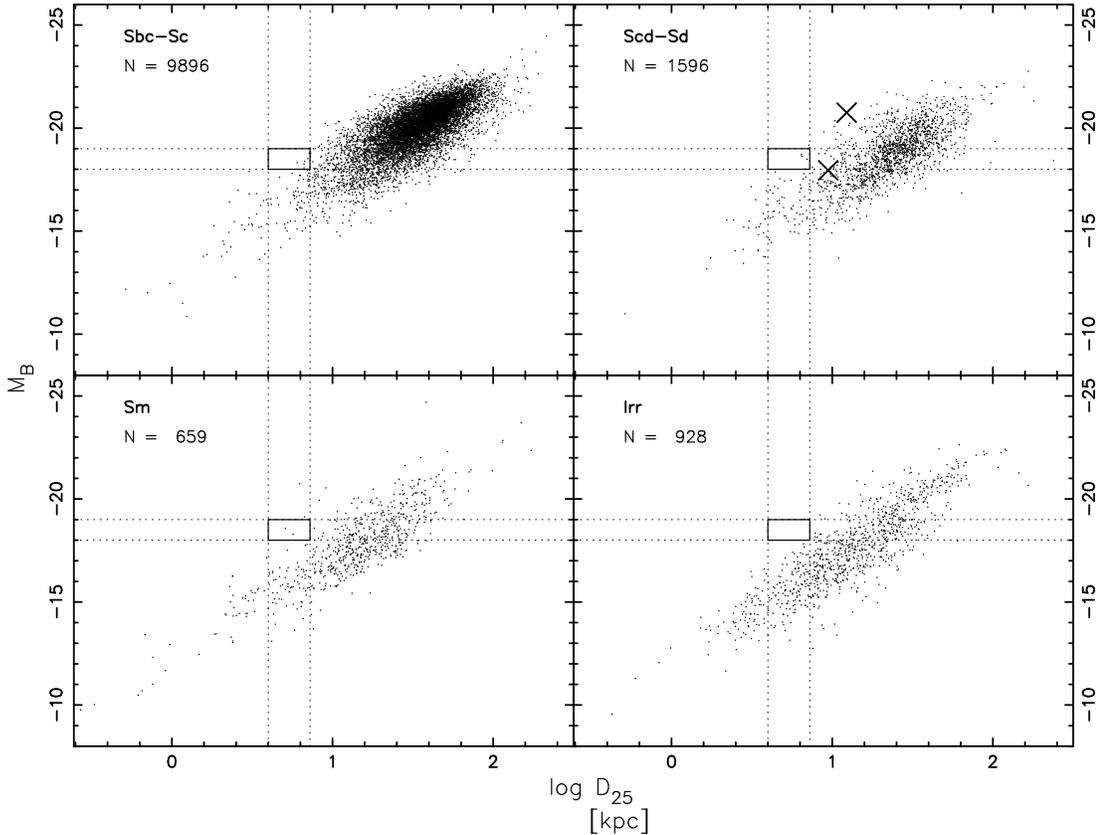}
  \caption{
  Absolute magnitude $M_B$ vs.\ linear diameter $\log D_{25}$
  diagrams for galaxy
  types Sbc-Sc, Scd-Sd, Sm, and Irr, as extracted from LEDA.  The region
  formed by the dotted lines, $-19.0 < M_B < -18.0$ and $4 < D_{25} < 7$
  gives the derived
  location for the GRB host
  galaxy in this diagram.  Crosses denote galaxies Mrk 1267 and NGC 7793 from
  Table 3. 
  Their continuum spectra are similar to that of the GRB host.
  }
  \label{fig6}
  \end{figure*}

  Figure \ref{fig6} shows the $M_B$ vs.\ $\log{D_{25}}$ diagrams for
 four ranges of galaxy
  types from LEDA (Sbc-Sc, Scd-Sd, Sm, Irr).  Different types have
    the well known relation, roughly following the dependence 
   $M \approx 5\log D +$ const.  The slope reflects what is expected from
    discs of different diameters with a constant average surface
    brightness.
  $M_B$ and $D_{25}$ are calculated as
  \[ M_B=B_{\rm T}-5\cdot \log{\frac{v}{H_0}}-25 \]
  \[ D_{25}= \frac{\pi}{108} \frac{v}{H_0} \cdot 10^{\log{d_{25}}} \]
  where $B_{\rm T}$ (total apparent B-magnitude),
  $\log{d_{25}}$ (log of 25 B-mag/arcsec$^2$ isophote diameter in units
  of $0.1$ arcmin), and $v$ (radial velocity in km s$^{-1}$)  are from LEDA.
  Isophotal diameters were corrected for inclination and galactic
  absorption, and radial velocities for Virgo infall.
    The horizontal dotted lines in Fig. 6
  indicate the allowed region of the present GRB host galaxy, as based on
  the absolute magnitudes derived from the models A, B and C.  

  In order to
  derive restrictions to the linear size, we use the 
 result from HST measurements of Fruchter et al. (1998): the
 GRB970508 host galaxy has an exponential disk with a scale length $r_0$ of
 about $0.07$ arcsec.

 For a model disk with an exponential distribution of surface brightness 
 from the center to galaxy periphery a simple relation between 
 the central surface brightness $\mu_0$, the 25 mag/arcsec$^2$ 
 isophotal angular radius $r_{25}$ (in arcsec), and   
 the scale length $r_0$ ($r_{25}/r_0$) 
  can be determined as:

 \[ 25-\mu_0 =  \frac{r_{25}}{r_0} 2.5 \log(e). \]
If for an exponential (model) disk one chooses as observational  
central surface brightness a value $\mu_0=20.4$ mag/arcsec$^2$, then
the fraction $r_{25}/r_0$ is $4.24$.
For the angular scale length $r_0=0.07$ arcsec,  
the corresponding angular diameter $2\,r_{25}$ turns out to be equal to $ \sim 0.6$ arcsec. 
The latter angular size of the GRB host galaxy corresponds to the isophote of 25 mag/arcsec$^2$
and is in good agreement with the upper limit of $\sim 0.4$ arcsec of the observable size of 
GRB host obtained in previous HST/NICMOS observations (Pian at al. 1997).
So, the linear diameter $D_{25}$ would be in the range of $\sim$ 4 -- 7 kpc 
for the A, B, C cosmological models (4) and   
the allowed region for $D_{25}$ is shown by vertical dotted lines in Fig. 6.

But in that case the GRB970508 host galaxy turns out to be in the region of blue compact
galaxies with heightened central surface brightness. 
In other words the galaxy is
on the same line in Fig. 6 (Scd--Sd) 
$M_B \approx -5\log D_{25} +$ const 
where the blue compact Mrk 1267   
with the continuum spectra (see Fig. 5) is. 
It good corresponds of the agreement (mentioned above, see Fig. 4) 
the continuum UV spectrum 
with the $BVR_cI_c$ broad-band GRB host spectrum.

  \section{Conclusions}

   The present late-time observations of the optical counterpart GRB970508 
   allows us to think that in 470 days after the GRB
   the observed $BVR_cI_c$ fluxes can be attributed only to the GRB host galaxy.
   The contribution of OT (if any at all) is already less than 
   observational errors. 
   It allows us to calculate  
   the intrinsic parameters of the GRB970508 host galaxy and to make 
   the first step in the study of the properties of the GRB host galaxy 
   by the comparison with the properties of galaxies in the local Universe, 
   on which we have good knowledge.

   But for the GRB redshift $z=0.835$, the derived absolute magnitude and 
   diameter,
   needed for comparison with local galaxies, depend on the
   adopted cosmological model.  A change in the model parameters easily
   produces a change of 1 mag in the absolute magnitude of the host galaxy
   and a change of $\sim$ 50\% in diameter. 
   We have measured the $K_B$ corrections for the GRB970508 host galaxy and the
   neighbouring G2 galaxy, which are + 0.6 and + 2.1, respectively.
   Such corrections make absolute magnitudes brighter and shift them away
   from dwarf galaxies.
   The position and luminosity of one of the neighbouring G2 
   allows it
   to produce the $z=0.767$ absorption line system in the spectrum of the
   GRB970508 OT on 11 May 1997. 

   The intrinsic physical properties of the possible
   host galaxy for GRB970508 correspond
   to
   a late-type galaxy, with its
   $M_B \approx -18.0$, $-19.0$ (depending on the cosmological model) clearly
   fainter than the knee of the general
   Schechter luminosity function at about $-20.8$ mag. However as demonstrated
    by Binggeli et al. (1988), the luminosity
   function essentially depends on the type of galaxies and their
   environment (field and clusters). For Sd + Sm galaxies the knee of
   the luminosity function is at $M_B \approx -18$ and hence the
   detected host of GRB970508 would be from the bright end for this
   galaxy type range.  In terms of the classification by Binggeli (1994),
   the host belongs to his Sequence 3 and lies well away from the border
   between
   normal and dwarf galaxies traditionally taken to be around $M_B = -16$.
   
   The spectral  distributions of normal galaxies with types earlier than Sbc are confidently
   excluded for GRB970508 host galaxy. 
   The comparison of the spectral  distribution of the host
   galaxy with local starburst galaxies revealed that the best fits were given
   by  the \HII\ galaxy NGC7793 and the blue
   compact galaxy Mrk 1267.
   Position of the host galaxy in the $M_B$ vs. $\log D_{25}$ diagram
   for local late-type LEDA galaxies and an analysis of the angular diameter 
   of its disc using the scale length of about $0.07$ arcsec from Fruchter
   et al. (1998) give evidence for a higher than normal surface brightness.
   The latter, together with the closeness of the spectral
   distribution for Mrk 1267 in the UV range of spectrum,
   allows us to attribute    
   GRB970508 host galaxy  
   to the blue compact galaxies.  

   Altogether, the observed absorption lines Mg\,{\sc ii},
   Mg\,{\sc i} and Fe\,{\sc ii}, the evidence for emission line [O\,{\sc ii}] and
   evidence for a high surface brightness
   are consistent with a starburst process going on in the present GRB host
   galaxy.

  \acknowledgements{ To A. I. Kopylov for the last observations in Aug. 1998.
 To T.N. Sokolova for the help in work with the text.
 We have made use of the Lyon-Meudon Extragalactic Database
  (LEDA) supplied by the LEDA team at the CRAL-Observatoire de Lyon (France).}
 This work has been partly supported by the Academy of Finland
 (project ``Cosmology in the local galaxy universe").
 Y.B. thanks for support by the Russian program ``Integration"
    project N.578.
The work was carried out under support of the ``Astronomy" Foundation
 (grant 97/1.2.6.4),
INTAS N 96-0315 and RBFI N98-02-16542.

 \end{document}